\documentclass[a4paper,11pt]{article}
\usepackage{pos}
\usepackage{slashed}

\newcommand{\GeV}{{\ensuremath\rm GeV}}
\newcommand{\TeV}{{\ensuremath\rm TeV}}
\newcommand{\fb}{{\ensuremath\rm fb}}
\newcommand{\pb}{{\ensuremath\rm pb}}

\title{LHC Benchmark scenarios in the TRSM}

\author*[a]{Tania Robens}

\affiliation[a]{Rudjer Boskovic Institute,\\
  Bijenicka cesta 54, 10000 Zagreb, Croatia}

\emailAdd{trobens@irb.hr}

\abstract{The TRSM is a new physics model that extends the scalar sector of the Standard Model by two additional CP even scalars. It leads to a large variety of interesting signatures, some of which have not yet been explored by the LHC experiments. I will also discuss the option to explore the hhh final state within this model.\\
 RBI-ThPhys-2022-45 }

\FullConference{%
  41st International Conference on High Energy physics - ICHEP2022\\
  6-13 July, 2022\\
  Bologna, Italy
}

\begin{document}
\maketitle
\section{The model}
The Two-Real-Singlet Model (TRSM) is a new physics model that enhances the Standard Model (SM) electroweak sector by two additional fields that are singlets under the SM gauge group and obey an additional $\mathbb{Z}_2\,\otimes\,\mathbb{Z}_2'$ symmetry. The model, including theoretical and experimental constraints, has been discussed in great detail in \cite{Robens:2019kga,Robens:2022nnw}. We assume that all scalar fields acquire a vacuum expectation value (vev) which softly break the above symmetry. As a consequence, all couplings to SM particles are inherited from the SM-like doublet via mixing. One of the three scalars has to have properties that comply with the measurements of the Higgs boson by the LHC experiments \cite{ATLAS:2022vkf,CMS:2022dwd}. At a hadron collider, typical production and decay processes are given by 
either asymmetric (AS) production and decay, in the form of
$p\,p\,\rightarrow\,h_3\,\rightarrow\,h_1\,h_2$,
where, depending on the kinematics, $h_2\,\rightarrow\,h_1\,h_1$ decays are also possible, or
symmetric (S) decays in the form of
$p\,p\,\rightarrow\,h_i\,\rightarrow\,h_j\,h_j$,
where in our study none of the scalars corresponds to the 125 \GeV~ resonance. Note that this in principle allows for further decays $h_j\,\rightarrow\,h_k\,h_k$, again depending on the specific benchmark plane (BP) kinematics.
We here use the convention that $M_1\,\leq\,M_2\,\leq\,M_3$ for the masses of the scalars $h_{1,2,3}$ in the mass eigenstates.
\section{LHC benchmark scenarios}
\begin{center}
\begin{figure}[htb!]
\begin{center}
\begin{minipage}{0.4\textwidth}
\begin{center}
\includegraphics[width=0.9\textwidth]{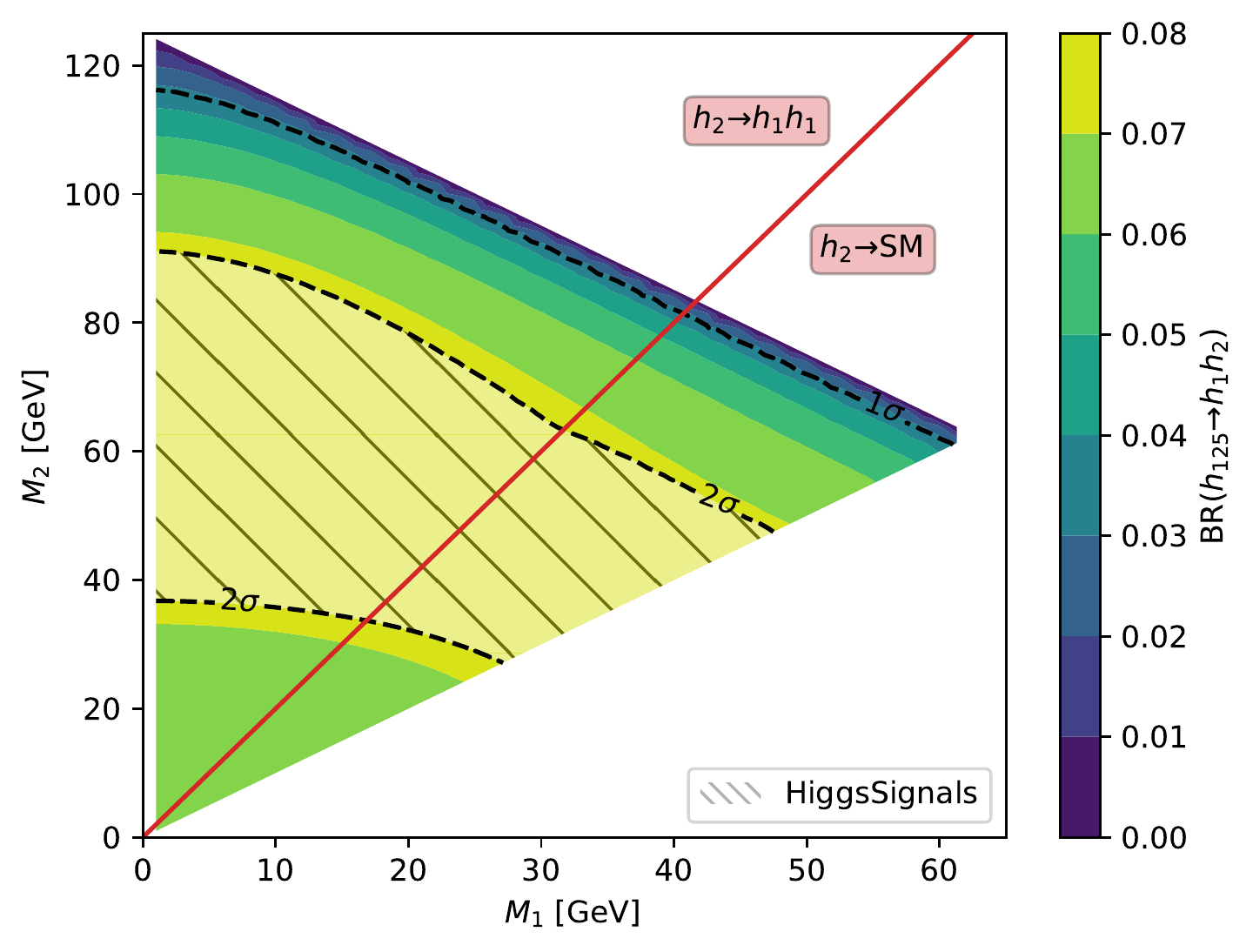}
\end{center}
\end{minipage}
\begin{minipage}{0.4\textwidth}
\begin{center}
\includegraphics[width=0.9\textwidth]{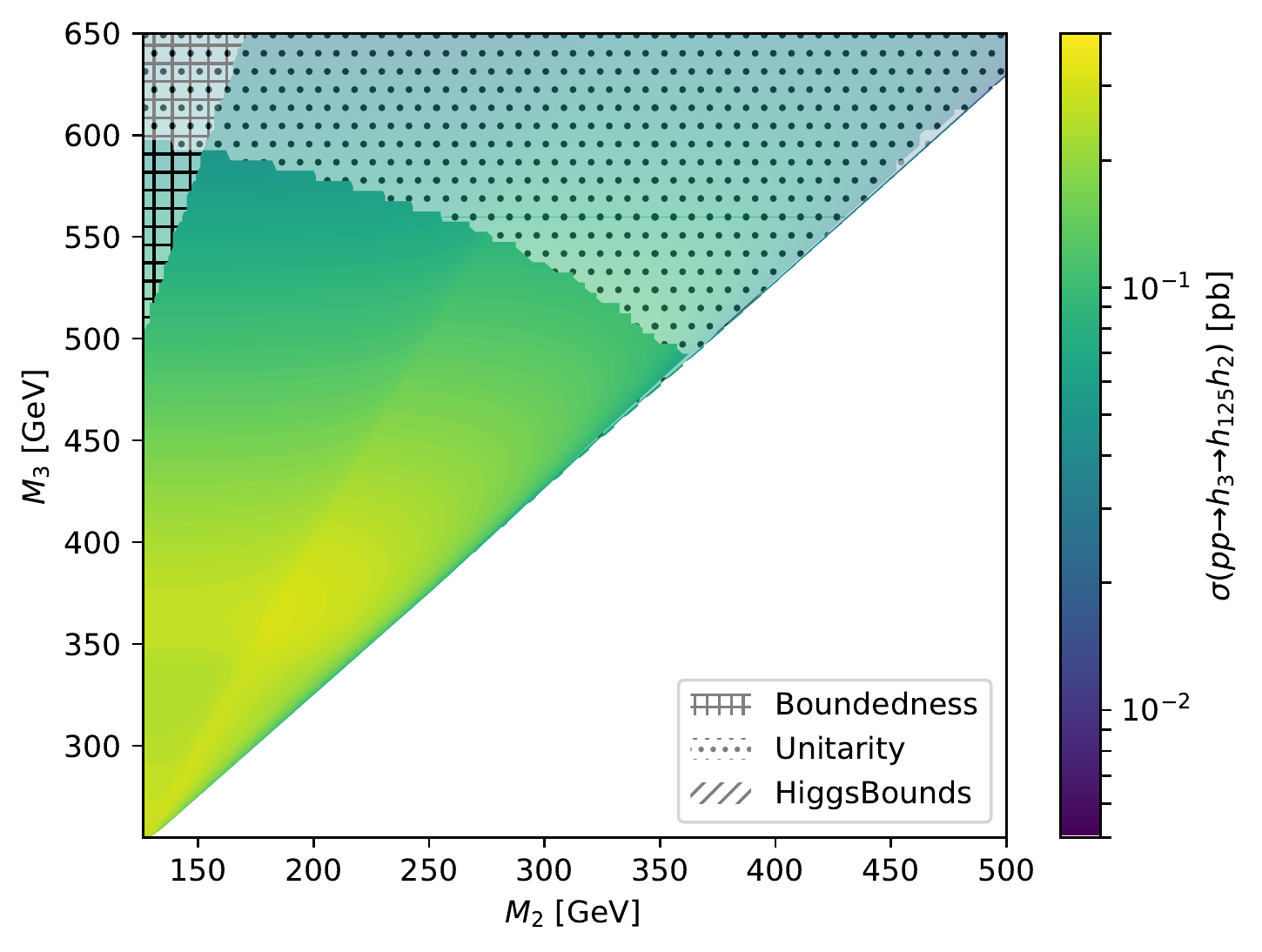}
\end{center}
\end{minipage}
\\
\begin{minipage}{0.4\textwidth}
\begin{center}
\includegraphics[width=0.9\textwidth]{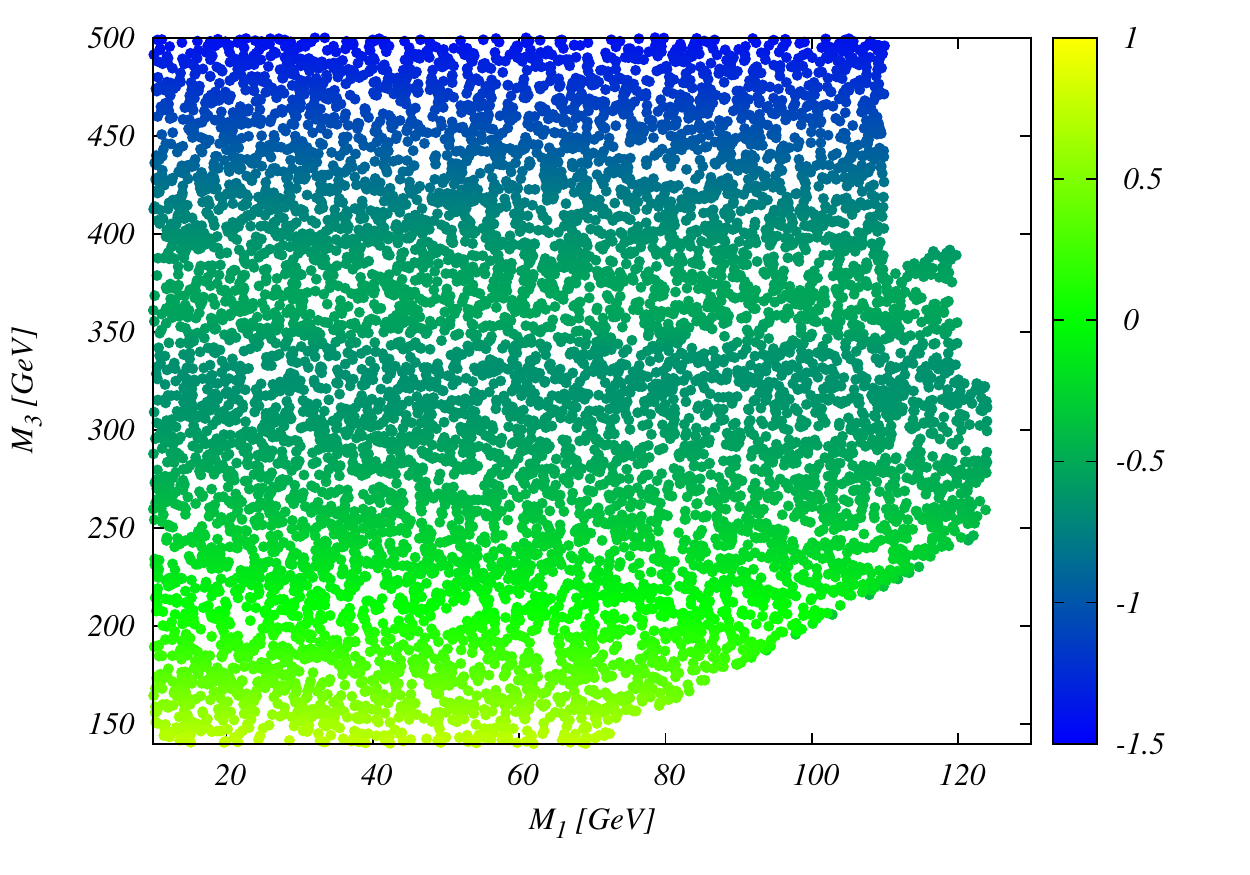}
\end{center}
\end{minipage}
\begin{minipage}{0.4\textwidth}
\begin{center}
\includegraphics[width=0.9\textwidth]{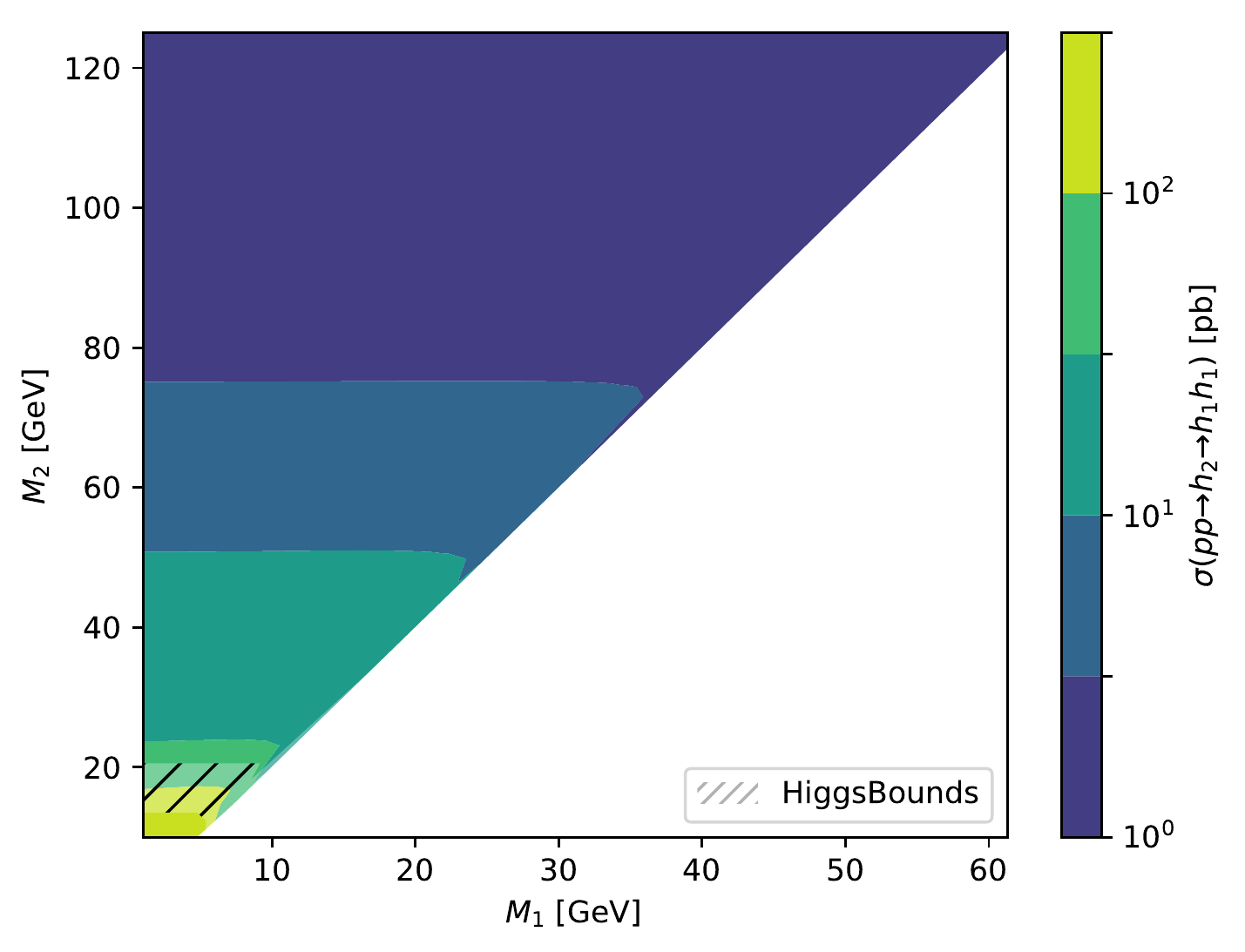}
\end{center}
\end{minipage}
\end{center}
\caption{\label{fig:bps} Benchmark planes in the TRSM.  {\sl Top:} Asymmetric final states, for $h_3\,\rightarrow\,h_1\,h_2$, BPs 1 {\sl (left)} and 3 {\sl (right)}. {\sl Bottom:} Symmetric final states, for $h_{i}\,\rightarrow\,h_j h_j$, where none of the $h$s has a mass of 125 \GeV, BPs 5 {\sl (left)} and 4 {\sl (right)}. Mostly taken from \cite{Robens:2019kga}, with an update for BP5 first presented in \cite{Robens:2022lkn,Robens:2022nnw}.}
\end{figure}
\end{center}

In \cite{Robens:2019kga}, several benchmark planes for both asymmetric and symmetric decay chains were proposed, which are distinct also by the role of the scalar that complies with the measurements of the 125 \GeV~ resonance. The benchmark planes, with maximal production cross sections at a 13 \TeV~ LHC, are given below.
\begin{itemize}
  \item {\bf   AS} {\bf   BP1: $h_3 \to h_1 h_2$ ($h_3 = h_{125}$) }: SM-like decays for both scalars: $\sim\,3\,\pb$; $h_1^3$ final states: $\sim 3\,\pb$
 \item{} {\bf   AS} {\bf   BP2: $h_3 \to h_1 h_2$ ($h_2 = h_{125}$)}: {SM-like decays for both scalars: $\sim\,0.6\,\pb$; $h_1^3$ final states: $\sim\,30\,\fb$
}
\item {\bf   AS} {\bf   BP3: $h_3 \to h_1 h_2$ ($h_1 = h_{125}$)}: { (a) SM-like decays for both scalars $\sim\,0.3\,\pb$; {  (b) $h_1^3$ final states: $\sim\,0.14\,\pb$}}
  \item {\bf   S} {}{\bf   BP4: $h_2 \to h_1 h_1$ ($h_3 = h_{125}$)}: {up to 60 \pb}
  \item {} {\bf   S} {\bf   BP5: $h_3 \to h_1 h_1$ ($h_2 = h_{125}$)}: { up to $2.5\,\pb$}
 \item {} {\bf   S} {\bf   BP6: $h_3 \to h_2 h_2$ ($h_1 = h_{125}$)}:
{SM-like decays: up to 0.5 \pb; {$h_1^4$ final states: around 14 \fb}}
\end{itemize}
We also show several of these benchmark planes in figure \ref{fig:bps}.
\section{hhh final states}
In BP3, an intriguing scenario is the one where also $h_2$ decays to $h_1 h_1$, which then leads to a $h_{125} h_{125} h_{125}$ final state. For the subsequent decay into $b\bar{b}$ pairs, this channel has been investigated in detail in \cite{Papaefstathiou:2020lyp}. 
 The results are shown in table \ref{tab:sigs}, where one sees that significances close to $5\,\sigma$ can already be achieved with relatively low integrated luminosity. 

\begin{center}
\begin{table}[h!]
\begin{center}
{\small
\begin{tabular}{c||cc||cc}\\
{\bf $(M_2, M_3)$}& $\sigma(pp\rightarrow h_1 h_1 h_1)$ &
$\sigma(pp\rightarrow 3 b \bar{b})$&$\text{sig}|_{300\rm{fb}^{-1}}$& $\text{sig}|_{3000\rm{fb}^{-1}}$\\
${[\GeV]}$ & ${[\fb]}$  & ${[\fb]}$ & &\\
\hline\hline
$(255, 504)$ & $32.40$ & $6.40$&$2.92$&{  $9.23$}\\
$(263, 455)$ & $50.36$ & $9.95$&{  $4.78$}&{  $15.11 $}\\
$(287, 502)$ & $39.61$ & $7.82$&{  $4.01$} &{  $12.68$}\\
$(290, 454)$ & $49.00$ & $9.68$&{  $5.02$}&{  $15.86 $}\\
$(320, 503)$ & $35.88$& $7.09$& {  $3.76 $}&{  $11.88$}\\
$(264, 504)$ & $37.67$ & $7.44$&{  $3.56 $}&{  $11.27 $}\\
$(280, 455)$& $51.00$ & $10.07$&{  $5.18$} &{  $16.39$}\\
$(300, 475)$&$43.92$& $8.68$&{  $4.64 $}&{  $14.68 $}\\
$(310, 500)$& $37.90$ & $7.49$&{  $4.09 $}&{  $12.94$}\\
$(280, 500)$& $40.26$& $7.95$&{  $4.00 $}&{  $12.65 $}\\
\end{tabular}}
\caption{\label{tab:sigs} Significances for the $h_{125} h_{125} h_{125}$ final states in the TRSM, at a 14 \TeV~ LHC, from the analysis presented in \cite{Papaefstathiou:2020lyp}, for various points in BP3.}
\end{center}
\end{table}
\end{center}

\section{Recasts}

\begin{center}
\begin{figure}[htb!]
\begin{center}
\begin{minipage}{0.45\textwidth}
\includegraphics[width=\textwidth]{bp5_new}
\end{minipage}
\begin{minipage}{0.45\textwidth}
\includegraphics[width=\textwidth]{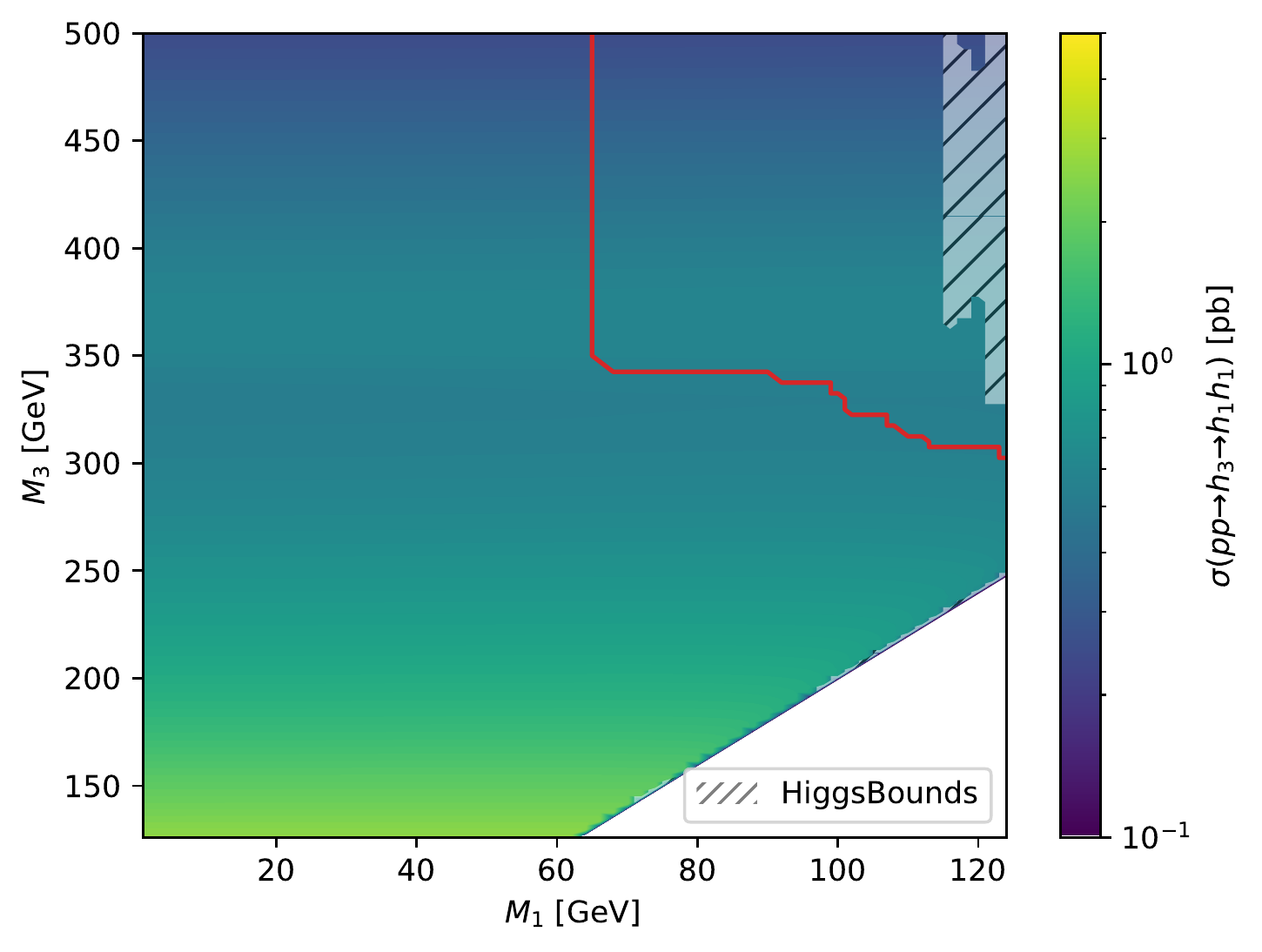}
\end{minipage}
\end{center}
\caption{\label{fig:recast} Recasting $h_{125}h_{125}$ search results from a $36\,\fb^{1-}$ CMS search \cite{CMS:2018qmt}, with recast results from \cite{Barducci:2019xkq}. Taken from \cite{Robens:2022mvi,Robens:2022lkn,Robens:2022nnw}.}
\end{figure}
\end{center}

It is also interesting to investigate whether current searches can be reinterpreted and recasted in such a way that they allow to exclude regions in the models parameter space that were not directly scrutinized in the experimental search.
 In \cite{Barducci:2019xkq}, the authors have reinterpreted a CMS search for $p\,p\,\rightarrow\,H\,\rightarrow\,h_{125}h_{125}\,\rightarrow\,4\,b$ \cite{CMS:2018qmt}. 
 I have applied these results to the TRSM, in particular to BP5. I display the corresponding results in figure \ref{fig:recast}. 
We see that the sensitive region of parameter space is significantly extended.
\section{Conclusion and outlook}
I discussed benchmark planes in the TRSM, a new physics model that enhances the SM scalar content by two additional neutral CP-even scalars that allows for interesting novel decay chains. I reported on benchmark planes, including a detailed study as well as current recast results. Althogh the experiments at the LHC already explore parts of the model's parameter space, there is still room for further investigation.

\end{document}